# A Multidatabase System as 4-Tiered Client-Server Distributed Heterogeneous Database System

Mohammad Ghulam Ali
Academic Post Graduate Studies and Research
Indian Institute of Technology, Kharagpur
Kharagpur, India
ali_iit@yahoo.com, ali@hijli.iitkgp.ernet.in

*Abstract*- In this paper, we describe a multidatabase system as 4-tiered Client-Server DBMS architectures. We discuss their functional components and provide an overview of their performance characteristics. The first component of this proposed system is a web-based interface or Graphical User Interface, which resides on top of the Client Application Program, the second component of the system is a client Application program running in an application server, which resides on top of the Global Database Management System, the third component of the system is a Global Database Management System and global schema of the multidatabase system server, which resides on top of the distributed heterogeneous local component database system servers, and the fourth component is remote heterogeneous local component database system servers. Transaction submitted from client interface to a multidatabase system server through an application server will be decomposed into a set of sub queries and will be executed at various remote heterogeneous local component database servers and also in case of information retrieval all sub queries will be composed and will get back results to the end users.

## I. INTRODUCTION

In recent years, multidatabase system has gaining attention of many researchers that attempts to logically integrate several different independent distributed heterogeneous DBMSs while allowing the local DBMSs to maintain complete control of their operations. A multidatabase system is a database system that resides on top of existing heterogeneous component local database systems and presents a single database illustration to its users [1,2]. The Multidatabase System usually maintains a single global database schema, which is integration of all local heterogeneous component databases schemas and against which users issue queries and updates. Multidatabase System maintains only global schema and the local component database system actually maintains all user data. Creating and maintaining the global schema, which requires the use of database integration techniques, is a critical issue in the multidatabase system.

Variety of approaches to schema integration have been proposed e.g. [3,4,5,6,9,10,11]. We are not considering this issue in this paper.

Keeping in mind the progress in communication and database technologies (concurrency, consistency and reliability) has increased the data processing potential.

Various protocols are proposed and implemented for network reliability, concurrency, atomicity, consistency, recovery and replication. The current demand is to access data from various existing heterogeneous database servers distributed among remote sites in a network and also to insert and update data at existing distributed heterogeneous databases which are autonomous and evolve over times. If any organization has headquartered in any country and has many branches across the globe, wants efficient and quick retrieval of information for any kind of decision supports.

In our proposed 4-tiered client-server architecture using a multidatabase system, a MDBS can be visualized as a client-server system that allows clients to simultaneously access and update data stored in more than one distributed database servers. **Layer 1** is a client Graphical User Interface or Web-based interface, which resides on the top of the Client Application Program or Application Server, **layer 2** is an application server that contains client program, business logic, API and access to Multidatabase system server. **Layer 3** is a Multidatabase System that controls and maintains Global Schema and Global Directory and access to various remote database servers based on user query. **Layer 4** is remote heterogeneous local component database servers.

A multidatabase is composed by a global schema and Global Database Management System. There is a Client Program, which resides on top of the Global Database Management System and Global Schema of the multidatabase system server. A global schema is created with a set of virtual global classes and is stored in a Global Database (GDB). User will submit a query on Global Schema through using web-based interface or graphical user interface of application program, the query will be decomposed into a set of sub-queries and will go to the respective remote local component databases servers and will be executed locally. The query will produce response. Incase of informational retrieval the sub-results coming from individual local component database servers are then will be composed and send back results to respective user.

In this proposed 4-tiered client-server architecture a client program on top of the multidatabase system servers manages and retrieves data from multiple sites within a single application through a multidatabase system server while providing complete autonomy to individual remote database systems. The multidatabase system server will maintain Global Directory contains metadata information of the global schema. This metadata can be queried locally to quickly obtain information about remote schemas.

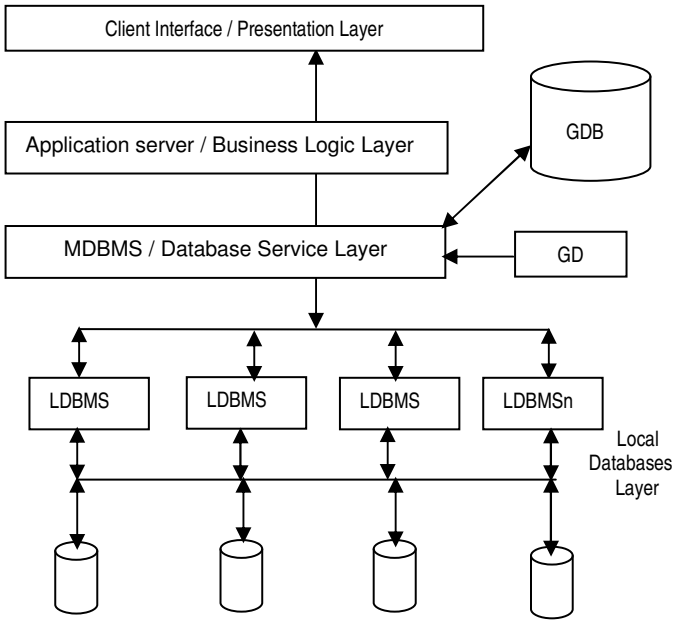

Figure 1. Overall View of 4-Tiered Client-Server Architecture

## II. INTEGRATION ISSUES OF HETEROGENOUS DISTRIBUTED DATABASES

Heterogeneous data source means there is no homogeneity among the databases at various sites or at least component databases differ in some important respect (e.g. the DBMS they are running or perhaps the data model implemented by it – relational, object-oriented, etc)

Following issues are to be considered during integration of heterogeneous distributed databases:

*A. Schema transformation.*

See details the work of [2,10,12]

*B. Correspondence Investigation.*

See details the work of [9,10,13]

- Semantic & Schematic Heterogeneity
- Semantic relevance/correspondence between classes

*C. Schema Integration and mapping.*

See details the work of [3,4,5,6,9,10,11].

*D. Schema evolution and automatic modifications propagation.*

See details the work of [1,10,14,15,16].

Many researchers have already addressed all these issues; we are not describing all these issues in this paper. However, we show two diagrams (Figure 2 and Figure 3) through which you can get overview of all these issues arise during integration of heterogeneous distributed local component databases and creation of a common data model and global schema.

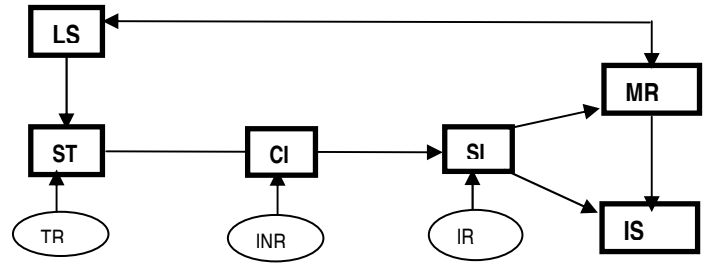

Figure 2. Generic framework for integrating heterogeneous local component database systems

Where

| | |
|---|---|
| LS | – is local schemas |
| ST | – is schema transformation |
| CI | – is correspondence investigation |
| SI | – is schema integration |
| MR | – is mapping rules |
| IS | – is integrated schema |
| TR | – is transformation rules |
| INR | – is investigation rules |
| IR | – Integration rules |

Using this techniques (Figure 2), the multidatabase system's designer usually transform the local schemes into a common data model, compare local schemas to exploit their semantic correspondences and identify their schematic differences, merge the local schemas into global schema (or a set of global schemas), and define the mappings between the global schema and the local schemas.

In the above approach as shown in Figure 2 and also in Figure 3 bellow, describes object-oriented data model as common data model to construct a global conceptual model by set a set of integration operators see details work of [2,9,10,11,12]. Since we have considered object-oriented data model as the data model of the multidatabase system in the proposed 4-tiered client-server architecture, the global schema as maintained in an object-oriented data model is a virtual schema because no actual data are stored for this schema. Thus, the classes in the global schema are called virtual classes and the objects associated with the classes are called virtual objects.

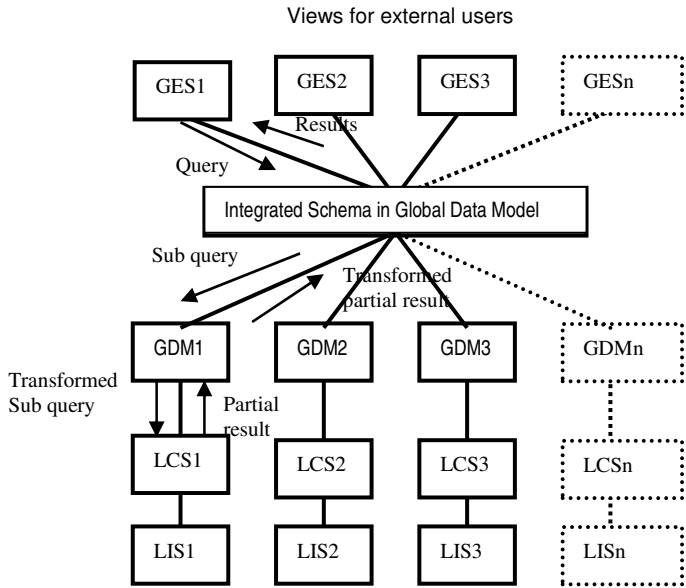

Figure 3. System Architecture of heterogeneous MDBS

Where

GDM – is local schema in Global Data Model
LCS – is logical organization at each site
LIS – is physical data organization on each site

### III. 4-TIERED CLIENT-SERVER ARCHITECTURE IN A MULTIDATABASE

This section describes a component-based reference architecture of a multidatabase system with a 4-tiered client-server architecture. We describe Multidatabase Structural Model and Multidatabase Logical Model.

*A. Structural Model*

The structural model as shown in figure 4 of the MDBS is a 4-tiered client-server model with a single multidatabase server. This model provides the advantages of a layered architecture, centralized control, global clients and distributed data access. As a disadvantage, this model introduces a single point of system failure. The four layers of the structural model are the global clients of the MDBS, the application server, the Multidatabase Management System Server (MDBMS) and heterogeneous distribute local component databases. **Layer 1** includes a software process capable of performing queries to the MDBMS server through application server. **Layer 2** holds the application server including application program, business logic and API. **Layer 3** is a Multidatabase System Server. **Layer 4** holds the server node of the MDBS. Layer 3 includes Multidatabase Management System (MDBMS), Global Schema and Global Directory. Layer 4 contains Local Database Management Systems (LDBMSs), local heterogeneous component databases that stores the distributed data.

If client interface is a Graphical User Interface (GUI), then application server at layer 2 will have client program and API. If client interface is a web-based interface, then the application server at layer 2 will have web server and web application. If client program is a Graphical User Interface then the client application will be published through thin-client on end users desktop.

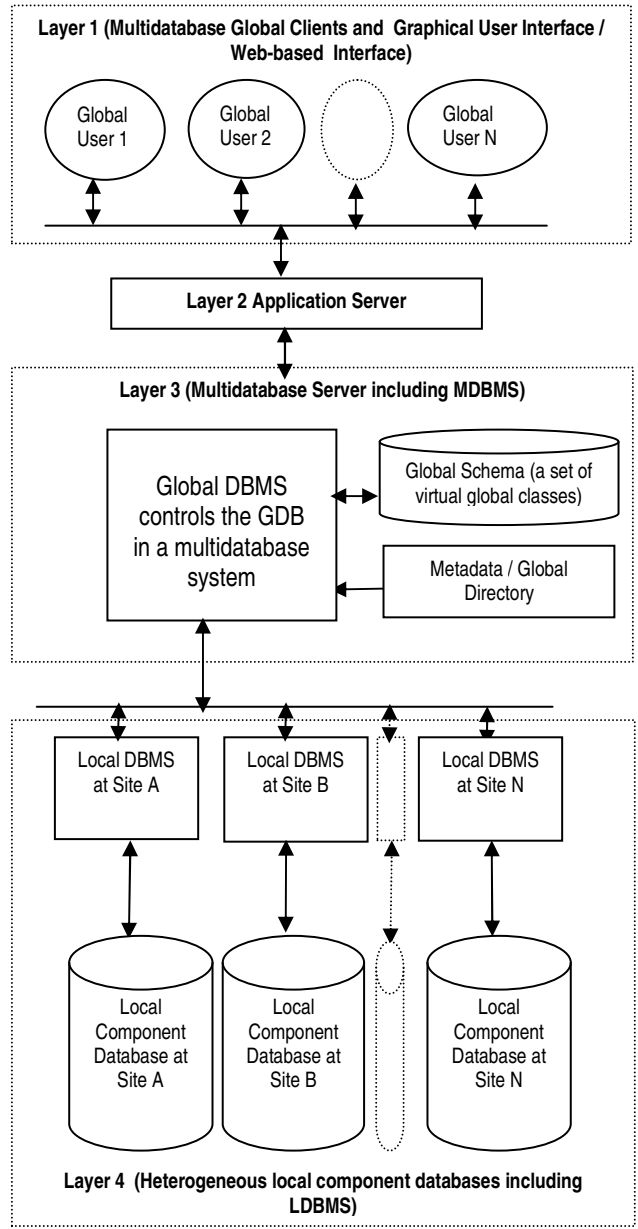

Figure 4. Structural Model

*B. Datalogical Model*

A classical example (as shown in Figure 5) of a data-based architecture is the ANSI/SPARC model by Tsichritzis and Klug [7].

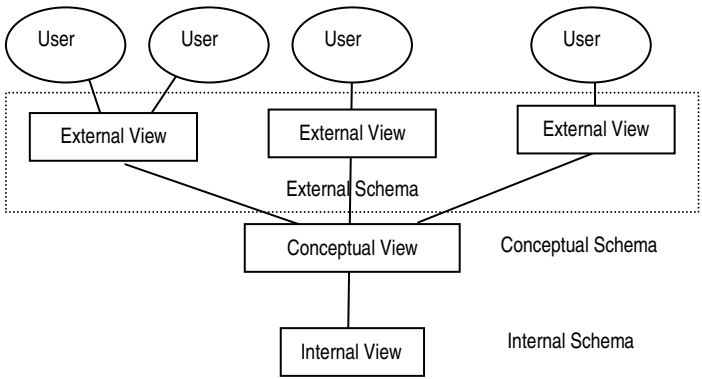

Figure 5. ANSI/SPARC Architecture

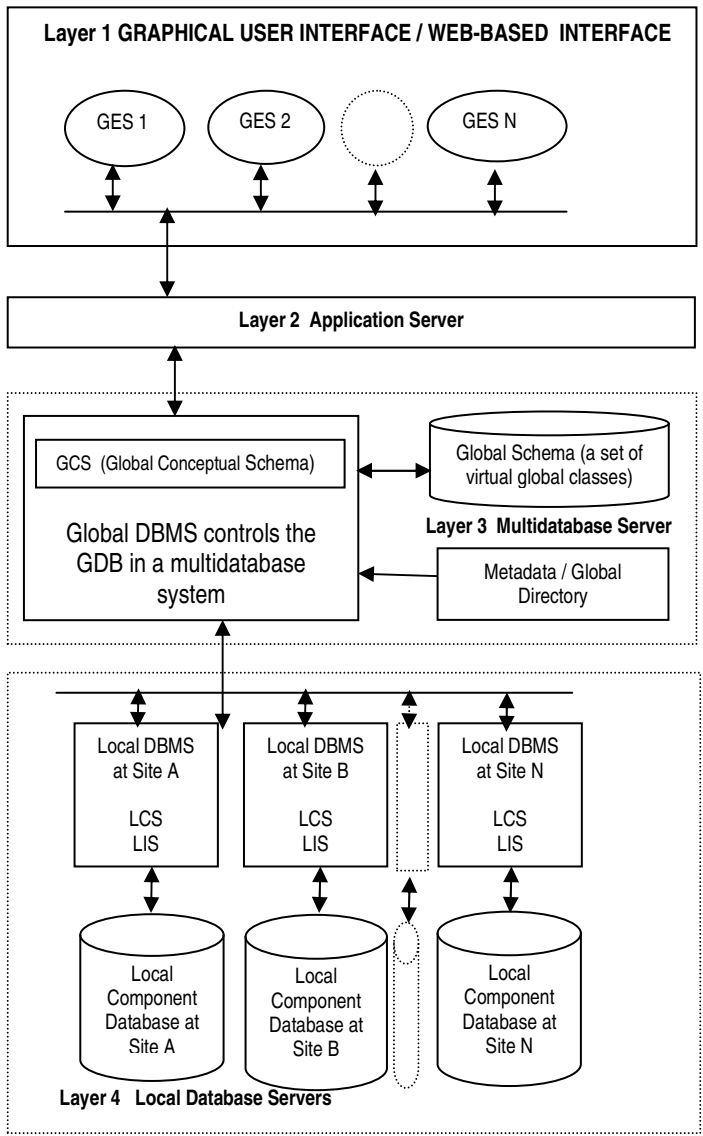

Figure 6. Datalogical Architecture

We adopt figure 5 in our proposed 4-tiered client-server model with single multidatabase server architecture model as shown in figure 6.

This model will contain the following components:

**GESn.** Views for external users. The n-th global external schema. This can be nth representation of databases through GUI or Web-based Interface.

**Application Server**. Stores client application program, business logic and API.

**MDBMS**. The database management system of the multidatabase. This include global conceptual schema. The information of these schemas is stored in the global directory and we say metadata information.

**GCS**. The Global conceptual schema of the multidatabase. The GCS describes the structure and constraints for the whole databases for community of users. We can say GCS is enterprise view of the database logical structure of the data at all sites. On the one hand, the heterogeneous database schemas have to be transformed into the global common data modal and on the other hand, the queries expressed in the global common data manipulation language have to be decomposed and translated into local query language.

**Metadata/Global Directory/Global Dictionary**. The Global directory/Metadata that stores both the global external schema and the global internal schema in order to permit the data transformation between these two schemas through mapping between their respective global definitions [8]. The MDBMS has global schema, the information about the GCS is held in global directory. This holds Meta data about what data objects exist at which site(s) in the MDBMS.

**LDBMSn**. The Database Management System (DBMS) of the n-th local heterogeneous component databases.

**Metadata/Local Directory/Local Dictionary**. Information about DB schema, relation, attributes, domain of attributes, relationship of DB.

**LCSn.** The local conceptual schema of the n-th local heterogeneous component database. LCS supports the logical organization at each site or abstract definition of the database.

**LISn.** The local internal schema of the n-th local databases. LIS supports the physical data organization on each site or deals with the physical definition and organization of the data.

**LDBn**. The nth local heterogeneous databases.

The Global Database Management System (GDBMS) will decompose query into a set of sub-queries including scanning, parsing and validating query. During this process, the system will use Global Directory/Metadata. Decomposed queries will pass to respective remote sites where Local Heterogeneous Component Database Management Systems are running. At each remote site, the query will be executed locally and during this process, the query will use metadata information of the remote site database. Now global and local query optimization is another issue in the multidatabase system. We are not taking into consideration this issue in this paper.

## IV. PERFORMANCE CHARACTERISTICS

In our proposed system, the 4-tiered client-server architecture will have following performance characteristics:

- Scalability
- Technological flexibility
- Lower long-term cost
- Better match business needs
- Improve customer services
- Reduce risk
- Code for business logic is centralized
- Client can not directly access remote database servers

## V. DATABASE SECURITIES

The Network security expert at each remote site can better protect remote site database server by implementing a firewall between the Global Database Server and the local database server and will examine each incoming packets coming to local database server, will authenticate this and will decide whether this packet is to be denied, dropped or forwarded to local database server. Since the IP address, port and the type of network service that the Global Database Management System is using in communicating with the remote database server is known by the firewall policy rules, can easily forward, drop or deny incoming packets. The DBA at each local site will provide a better database server level and database object level security. The System Administrator at each local site will provide a better OS level security. How to exactly tackle all these issues, we are not explaining in details in this paper.

## VI. CONCLUSION

This paper proposes reference architecture of a multidatabase system with a 4-tiered client-server structure model. The main objective of the work is to provide transparent access to autonomous, and distributed relational databases. This is a viable proposed system in the integration of Multidatabase System, implementation of 4-tiered client-server architecture and easy to maintain the Global Schema. In future, we plan to address other issues involving multidatabase system as 4-tiered client-server architecture.

AUTHORS PROFILE

**Mohammad G. Ali** I am a System Engineer Grade I in the Indian Institute of Technology Kharagpur, West Bengal, India. I am associated with System Analysis & Design, Programming, Implementation and Maintenance of Client-Server DBMSs and Web Application Developments. I am also associated with Database Administration, Web Server Administration, System Administration and Networking of the Institute. I have deployed many small to big projects on our Institute Network.